\documentclass[prb,preprint,citeautoscript]{revtex4-1} 
\usepackage{amsfonts}
\usepackage{amssymb}
\usepackage{amsmath}
\usepackage{xcolor}
\usepackage{esint}
\usepackage{natbib}
\usepackage[english]{babel}
\usepackage{physics}
\usepackage{siunitx}
\usepackage{graphicx}
\DeclareSIUnit\year{year}
\newcommand{\veff}{v_{\text{eff}}}
\usepackage{hyperref}
\hypersetup{colorlinks=true,
citecolor=black,
filecolor=black,
linkcolor=black,
urlcolor=black}
\begin{document}
\title{Application of the heat equation to the study of underground temperature}
 \author{CAPRAIS Mathis}
 \email[Correspondence email address: ]{mathis.caprais@universite-paris-saclay.fr}
 \affiliation{Université Paris-Saclay, 91190, Gif-sur-Yvette, France.}
 \author{PENSEC Ugo}
 \affiliation{Sorbonne Université, CNRS/IN2P3, Laboratoire de Physique Nucléaire et de Hautes Énergies, France}
 \author{SHVIRO Oriane}
 \affiliation{Université Paris-Saclay, 91190, Gif-sur-Yvette, France.}
 \author{ZEYEN Hermann}
 \affiliation{Département des Sciences de la Terre, UMR 8146, Université Paris-Saclay, CNRS, 91405 Orsay, France}
\date{\today}
\begin{abstract}
Modeling underground temperatures provides a practical application of the one-dimensional heat equation. In this work, the one-dimensional heat equation in surface soil is extended to include heat carried by the vertical flow of rainwater through the soil. Analytical solutions, with and without water flow, illustrate the influence of rainwater circulation on the sub-surface propagation of seasonal temperature variations, an important effect that is generally neglected in textbooks. The surface temperature variations are damped by the soil, and this effect was used by the Troglodytae in Egypt or the Petra in South Jordan to insulate against extreme temperatures. For a realistic case of horizontally layered geology, a finite difference Python code was developed for  the same purpose. Subsurface temperatures were also measured over a full year at depths up  to 1.8 m and used to estimate the thermal skin depth and thermal wavelength. This study  provides students with a practical example of how a textbook physics problem can be modified to extract information of contemporary importance in geophysics and global warming.
\end{abstract}
\maketitle
\section*{Introduction}
The heat equation describes how energy is transferred from one body to another as the result of a temperature gradient. It describes a wide variety of phenomena over a broad range of size scales, from heat diffusion in nanoscale materials to the macroscopic motion of atmospheric air. It is used to model simple, but important problems such as the insulation of a house, a furnace or a thermos flask. It is also the simplest diffusion equation, a class of fundamental expressions that includes Schrödinger's equation.

This study was initiated by an interest in underground buildings as energy-efficient constructions \cite{alkaff2016review}. Below a certain depth, the near-surface soil has a temperature that is almost constant throughout the year \cite{robson1989elementary}. Combined with the large thermal inertia of the underground environment, this allows for significant cooling and heating energy savings. This paper investigates both theoretically, numerically and experimentally how temperature is distributed below the Earth's surface, demonstrating the concept of a thermal skin depth. This is a classic undergraduate problem that is solved in several textbooks \cite{robson1989elementary}. However, textbooks as well as recent publications in geophysics usually neglect the role of heat advection induced by rainwater flowing vertically through the soil \cite{xing2017prediction1}. The introduction of rainwater advection transforms a classic  textbook problem into one of contemporary importance and interest.
\section{Theoretical framework}
\subsection{Derivation of diffusion equation without and with water advection}\label{intro_heat_eq}
Fourier's Law describes how heat propagates in a solid medium \cite{fourier1808memoire}:
\begin{equation}
\vb{q}_{d}=-\lambda \grad{T},
\label{eq:fourier}
\end{equation}
where $\vb{q}_d$ is the diffusive heat flux, $\lambda$ is the thermal conductivity of the medium, and $T$ is the temperature. For an incompressible medium at constant pressure, no work is done and heat flow is equal to the change in internal energy, or
\begin{equation}
    \pdv{}{t}\qty(\rho c T) = - \div{\vb{q}_d} = \div{\lambda \grad{T}},
    \label{eq:heat_eq}
\end{equation}
where $\lambda$ can be a function of space and time, and $\rho$ and $c$ are the local density and specific heat, respectively. When the medium is porous, a fluid -- in this case rainwater -- can flow through it and transfer additional heat. This advective heat flux is expressed as:
\begin{equation}
\vb{q}_{a}=\rho_w c_{w} T \varphi \vb{v},
\label{eq:advective}
\end{equation}
where $\rho_w$ is the fluid density, $c_w$ its specific heat at constant pressure, $\vb{v}$ its velocity, and $\varphi$ is the porosity of the medium. The quantity $\varphi \vb{v}$ is called the Darcy velocity, and describes the effective water transfer per unit area through the porous medium. To account for the advective heat flux, Eq. \eqref{eq:heat_eq} is easily generalized:
\begin{equation}
    \pdv{}{t}\qty(\rho c T) = - \div{\qty(\vb{q}_d + \vb{q}_{a})}.
    \label{eq:heat_eq_advection_diffusion}
\end{equation}
For our purposes, the medium is considered to be a collection of horizontal layers, that are constant with respect to time and independent of horizontal position. The fluid flow is assumed to be strictly vertical (Fig. \ref{coupe_sol}). In this case, Eq. \eqref{eq:heat_eq_advection_diffusion} can be written as:
\begin{equation}
\rho c\pdv{T}{t}=\pdv{}{z}\lambda\pdv{T}{z}-\rho_w c_w\varphi v \pdv{T}{z},
\label{eq:heat_equation_incompressible_full}
\end{equation}
where $\lambda_t$ is the weighted thermal conductivity. To find analytical solutions, the layered medium is homogenized in Sec. \ref{sec:analytical}. New quantities are introduced to simplify Eq. \eqref{eq:heat_equation_incompressible_full}: $\rho_t$, $c_t$ and $\lambda_t$ which are the weighted density, the weighted heat capacity and the weighted thermal conductivity, respectively. For the homogenized system, Eq. \eqref{eq:heat_equation_incompressible_full} reduces to,
\begin{equation}
\rho_t c_t\pdv{T}{t}=\lambda_t\pdv[2]{T}{z}-\rho_w c_w\varphi v \pdv{T}{z}.
\label{eq:heat_equation_incompressible}
\end{equation}
When the fluid is static ($\vb{v}=\vb{0}$), Eq. \eqref{eq:heat_equation_incompressible} can be written as a simple diffusion equation:
\begin{equation}
\pdv{T}{t} = D \pdv[2]{T}{z},
\label{reminder_heat_equation}
\end{equation}
where $D=\frac{\lambda_t}{\rho_t c_{t}}$ is the thermal diffusivity.
\subsection{General solutions for both cases}\label{intr:th_frame_eq}
For a static flow, ($\vb{v}=\vb{0}$), Eq. \eqref{reminder_heat_equation} can be solved by writing the temperature distribution as a Fourier series\cite{sabieh2014fourier}. Assume that the temperature at $z=0$, the Earth's surface temperature, is known at all times and can be written as a Fourier series:
\begin{equation}
T(0,t)=C_0 + \sum^{+\infty}_{n=1} C_n \cos \left( \omega_n t -\Delta_n\right),
\label{reminder_initial_conditions}
\end{equation}
with $C_0$ the long-term average of the surface temperature. $C_n$ is the amplitude of the temperature variation with period $\tau_n = 2 \pi / \omega_n$ and $\Delta_n$ is the phase angle of the $n^{th}$ harmonic, depending on the choice of the time origin. The boundary condition at $z=0$ reflects the daily and seasonal modulations of the surface temperature. For a $z$-axis oriented downwards, it is the solution mentioned by Kusuda and Achenbach \cite{kusuda1965earth, carslaw1959conduction}:
\begin{equation}
T(z,t)=C_0+\sum_{n=1}^{+\infty} C_n \exp(- \frac{z}{\delta_n}) \cos(\omega_n t-\Delta_n- \frac{z}{\delta_n}) \qq{with} \sqrt{\frac{\pi}{D\tau_n}}=\frac{1}{\delta_n},
\label{reminder_full_sol}
\end{equation}
It should be noted that this solution is not valid for very large depths as the decay of radioactive isotopes becomes an important heat source. Eq. \eqref{eq:heat_equation_incompressible} incorporates water flow effects,
\begin{equation}
\pdv{T}{t}=D\pdv[2]{T}{z}-\veff \pdv{T}{z},
\label{heat_equation_mod}
\end{equation}
where $\veff$ is an effective water speed, 
\begin{equation}
\veff=v\frac{\rho_w c_{w}\varphi}{\rho_{t} c_{t}}.
\label{eq:effective_speed}
\end{equation}
While the Darcy velocity is independent of $z$, $\veff$ is not necessarily constant, because both the heat capacity and the density can have a spatial dependence. Similarly to the above treatment, the general solution can be decomposed into time harmonics:
\begin{equation}
T_n(z,t)=C_n \exp(i\qty(\omega_{n} t-k^\prime_{n}z-\Delta_n)-k_{n}z),
\label{complex_sol}
\end{equation}
where $k_{n}$ and $k^\prime_{n}$ are real numbers, and $i=\sqrt{-1}$. In Eq. \eqref{complex_sol}, $k_{n}$ is an attenuation factor and $k^\prime_{n}$ a phase factor. The solution has to satisfy,
\begin{equation}
T(z=0,t)=C_0+\sum_{n=1}^{+\infty} C_n \exp(i\omega_{n} t-\Delta_n),
\label{eq:initial_conditions}
\end{equation}
and should be bounded for $z\to \infty$ and $k_n \geq 0$. Inserting Eq. \eqref{complex_sol} in Eq. \eqref{heat_equation_mod}:
\begin{equation}
i\omega_{n} C_n=D\left(k_{n}+ik^\prime_{n}\right)^{2}C_n+\veff\left(k_{n}+ik^\prime_{n}\right)C_n.
\label{injection}
\end{equation}
Equating the real and imaginary parts of Eq. \eqref{injection} separately gives:
\begin{equation}
   \begin{cases}
      & 0  = D\left( k_{n}^{2} - k_{n}^{\prime\,2}\right)+\veff k_{n}, \\
      & \omega_{n} = 2 Dk_{n}k^\prime_{n}+\veff k^\prime_{n}.
   \end{cases}
   \label{real_part}
\end{equation}
From the second equation, the following is obtained:
\begin{equation}
k^\prime_{n}=\frac{\omega_{n}}{2 Dk_{n}+\veff}.
\label{subs}
\end{equation}
Finding the solutions for $k_n$ then amounts to solving the polynomial equation:
\begin{equation}
D\left(2 Dk_{n}+\veff\right)^{2} k_{n}^{2}+\veff\left(2 Dk_{n}+\veff\right)^{2}k_{n}-D\omega_{n}^{2}=0.
\label{polynom}
\end{equation}
Only the positive real solution is retained from this:
\begin{equation}
k_{n}=\frac{1}{\delta_n}=\frac{-\veff\sqrt{2}+ \sqrt{\veff^{2}+\sqrt{\veff^{4}+16 D^{2}\omega_{n}^{2}}}}{2\sqrt{2} D},
\label{sec_roots}
\end{equation}
where $\delta_n$ represents the thermal skin-depth beyond which the surface temperature variation is damped  by a factor larger than $e$. One can also define the thermal wavelength:
\begin{equation}
\Phi_n=\frac{2\pi}{k^\prime_n}=2\pi \qty(\frac{\frac{2D}{\delta_n}+\veff}{\omega_n}).
\label{phase}
\end{equation}
The thermal wavelength is the spatial period of the underground thermal oscillations. From Eq. \eqref{phase}, it is an increasing function of the effective velocity $\veff$ as the temperature oscillation is carried by the water flow. The solution of Eq. \eqref{eq:heat_equation_incompressible} can finally be written in the general form:
\begin{equation}
T(z,t)=\sum_{n=0}^{+\infty} C_{n}\exp(\frac{-z}{\delta_{n}})\cos \qty(\omega_{n}t-\frac{2\pi z}{\Phi_{n}}-\Delta_n).
\label{final_solution}
\end{equation}
When taking $\veff = 0$ in Eq. \eqref{final_solution}, the temperature distribution reduces to known Kusuda-Achenbach solution \cite{kusuda1965earth, carslaw1959conduction}. Indeed, the thermal skin-depth $\delta_n$ becomes directly proportional to the thermal wavelength $\Phi_n$.

In the remainder of this paper, only $n=0,1,2$ are considered as these harmonics corresponds to the dominant timescales of the surface temperature variations. $n=0$ corresponds to the long-term average temperature, $n=1$ to the daily cycle and $n=2$ to the annual cycle. Then, $C_0 = T_0$ with $T_0$ the average temperature throughout the year, $C_1 = T_{1}$ the amplitude of the temperature variation during the day-night cycle, and $C_2 = T_2$ the temperature variation corresponding to the seasonal cycle. The angular frequency of the day-night cycle is $\omega_{1}$, $\omega_2$ the angular frequency of one year and $\Delta_1$ and $\Delta_2$ are phase terms.

With this approximation, the thermal skin-depth is of the order of \SI{10}{\meter}, which is the depth at which the analytical study is conducted in the following section.
\section{Analytical results for averaged properties}\label{sec:analytical}
The above results will now be applied to a real situation: the ground temperature of the Plateau de Saclay, a geological formation located near Paris, France. As will be shown, the skin depths at this location satisfy $\delta_1 < \delta_2 < \SI{6}{\meter}$, so the analysis is restricted to depths $z < \SI{10}{\meter}$. In Sec. \ref{subsec:geo}, the averaged thermal properties of the Plateau are calculated. In Sec. \ref{subsec:skin_depth}, the thermal skin-depth and thermal wavelength are evaluated for the Plateau de Saclay and the effect of water flow is discussed.
\subsection{Averaged thermal properties of 5-layer system}\label{subsec:geo}
\begin{figure}[hbtp]
    \centering
    \includegraphics[width=120mm]{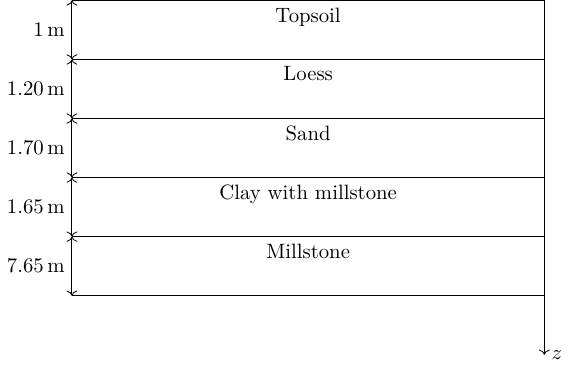}
    \caption{Geologic cross-section of Plateau de Saclay.}
    \label{coupe_sol}
\end{figure}

Geological studies\cite{vernoux1999etude} show that the Plateau has five distinct strata within this depth (Fig. \ref{coupe_sol}). The physical properties of each layer are summarized in Table \ref{tab:properties}. To obtain an analytical solution, we treat the five strata as a single homogeneous layer with physical properties determined by weighted
averaging over the properties of the five strata. To calculate the ground temperature vs. depth and time, the values of $\lambda$ and $\rho c$ are required. For a layered structure, the average thermal conductivity is given by\cite{carslaw1959conduction}:
\begin{equation}
    \frac{1}{\lambda_t}=\sum_{i} \frac{\alpha_{i}}{\lambda_{i}},
    \label{eq:thermal_conductivity}
\end{equation}
(akin to the addition of resistances in series), where the coefficients $\alpha_i$ represent the volume fraction of the $i$-th layer in the five-layer system ($\sum_i \alpha_i = 1$). The total volumetric heat capacity is defined as the arithmetic average of the various heat capacities,
\begin{equation}
    c_{t}\rho_t=\sum_{i} c_{i} \rho_i \alpha_{i}.
    \label{eq:heat_capacity}
\end{equation}
\begin{table}[hbtp]
\centering
\begin{tabular}{cccc}
\hline
Properties & $\rho c$ (\SI{}{\joule\per\kelvin\per\cubic\meter}) & $\lambda$ (\SI{}{\watt\per\meter\per\kelvin}) & $D$ (\SI{}{\square\meter\per\second}) \\
\hline
\hline
Topsoil\cite{zaibon2019soil} & $\SI{3431\pm7e3}{}$ & $\SI{1.38\pm0.01}{}$ & $\SI{402\pm 3e-9}{}$ \\
Loess\cite{zhen2019thermal,article_loess} & $\SI{225\pm2e4}{}$ & $\SI{1.5\pm 0.1}{}$ & $\SI{66\pm 4e-8}{}$ \\ 
Sand\cite{deepagoda2016characterization,hamdhan2010determination} & $\SI{3200\pm6e3}{}$  & $\SI{2.0\pm0.4}{}$ & $\SI{6\pm1e-7}{}$ \\
Clay with millstone\cite{cote2005generalized,geothermie} & $\SI{2331\pm5e3}{}$  & $\SI{1.3\pm0.2}{}$ & $\SI{56\pm9e-8}{}$ \\
Millstone \cite{manger1963porosity,orlander2018thermal,abid2014thermophysical} & $\SI{2435\pm5e3}{}$ & $\SI{5.1\pm0.5}{}$ & $\SI{21\pm 3e-8}{}$ \\
Water\cite{alma,ginnings1953heat,ramires1995standard} & $\SI{417\pm3e4}{}$  & $\SI{0.600\pm 0.003}{}$ & $\SI{144\pm 1e-9}{}$ \\
Total & \SI{2.3\pm0.2e6}{} & \SI{2.3\pm0.5}{} & \SI{1\pm 0.2 e-6}{} \\
\hline
\end{tabular}
\caption{Soil properties for a saturated soil, where $D = \lambda / \rho c$. 
For the clay with millstone layer, in the absence of data in literature, it was assumed that \(\rho\) and \(\lambda\) were those of gritstone, while the heat capacity was taken to be that of sandstone (ref. \onlinecite{cote2005generalized,geothermie}).}
\label{tab:properties}
\end{table}
\subsection{Evaluation of the thermal skin-depth and thermal wavelength}\label{subsec:skin_depth}
In the absence of water flow ($\vb{v} = \vb{0}$), using Eqs. \eqref{eq:thermal_conductivity} and \eqref{eq:heat_capacity} to compute $D$ reported in Table \ref{tab:properties} the skin depths $\delta$ are found to be \(\delta_1 = \SI{0.17\pm 0.01}{\meter}\) and \(\delta_2=\SI{3.1\pm 0.5}{\meter}\). The thermal wavelengths, calculated from Eq. \eqref{phase}, are \(\Phi_{1}=\SI{1.0\pm0.1}{\meter}\) and \(\Phi_2 = \SI{19\pm2}{\meter}\). A larger layer thickness is needed to dampen the annual temperature variations than the daily ones.

More realistically, one must include heat transport due to water flowing through the soil. For the Plateau de Saclay, the Darcy velocity has been measured\cite{vernoux1999etude} to be $v\varphi=\SI{2.2e-7}{\meter\per\second}$. In the following, $v$ and $v\varphi$ are assumed to be constant throughout the year. While this implies continuous and constant precipitation throughout the year, the rainfall near the Plateau nearly satisfies this assumption. From Eq. \eqref{eq:effective_speed} and Table \ref{tab:properties}, $\veff=\SI{3.9\pm0.2e-7}{\meter\per\second}$, and using Eqs. \eqref{sec_roots} and \eqref{phase}, \(\delta_1 = \SI{0.17\pm 0.02}{\meter}\), \(\delta_{2}=\SI{6.7\pm 0.5}{\meter}\), \(\Phi_{1}=\SI{1.0\pm0.1}{\meter}\) and \(\Phi_{2}=\SI{22\pm 2}{\meter}\). Note that $\delta_2$ with water flow is almost twice its value without water flow. Since the annual temperature
variation greatly exceeds the diurnal variation, and $\delta_2 \gg \delta_1$. With or without water flow, as the magnitude of the yearly skin depth $\delta_2$ is much larger than the daily skin depth, the daily variations are not be studied in the remainder of this paper.
\begin{figure}[hbtp]
    \centering
    \includegraphics[width=120mm]{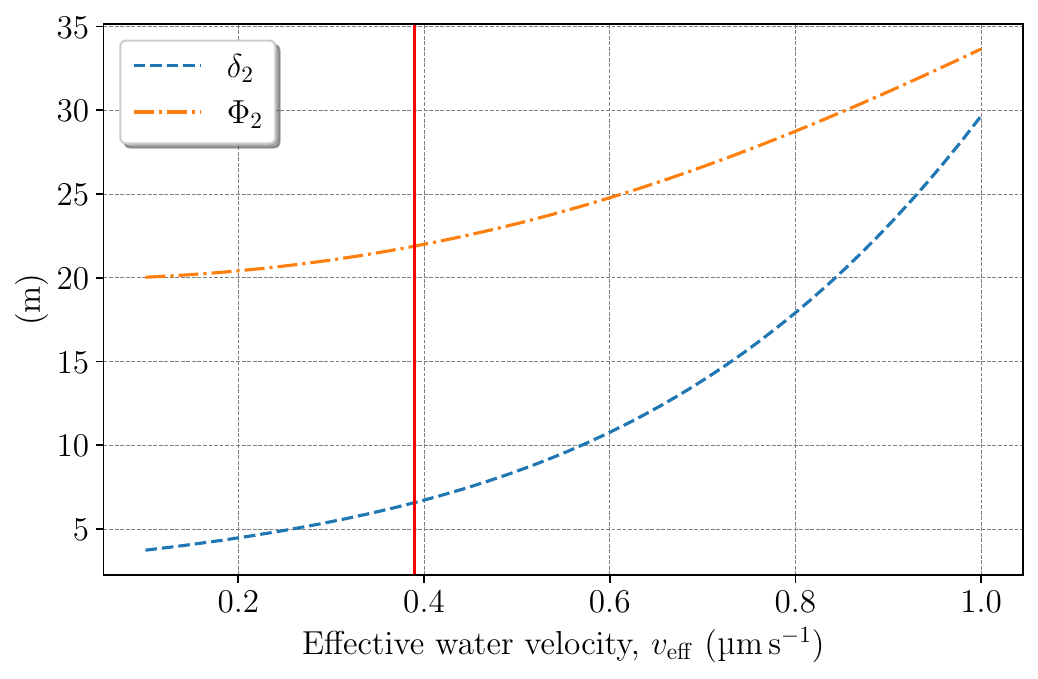}
    \caption{Plots of $\delta_2$ and $\phi_2$ (annual period) as a function of the underground effective speed of water (Eqs. \eqref{sec_roots} and \eqref{subs}). The red line represents the velocity of the water flow used in the modelling.}
    \label{both_sol}
\end{figure}
In Fig. \ref{both_sol}, the evolution of the annual thermal skin depth (\(\delta_2\)) and the annual thermal wavelength (\(\Phi_2\)) for typical velocities of water circulation in sediments is represented. For small velocities, $\Phi_2$ shows less sensitivity to variations of the effective water speed than $\delta_2$. As expected, with an increase in water flow, both \(\delta\) and \(\Phi\) rapidly increase. This is due to water flow carrying thermal perturbations throughout the soil. The vertical line in the graph represents the velocity used in both the analytical solution and the numerical simulations.
\section{Numerical solution}\label{sec:numerical}
The analysis conducted in Sec. \ref{sec:analytical} considered with a single homogeneous layer whose properties were obtained from the five-layer system. To account for the spatial variation of the heat capacity, the thermal conductivity and effective velocity, it is necessary to solve Eq. \eqref{eq:heat_equation_incompressible_full}. The discretized problem is presented in Sec. \ref{subsec:discretization}. The results for the homogeneous medium are presented in Sec. \ref{subsec:results_homogeneous}, and for the heterogeneous medium in Sec. \ref{subsec:results_heterogeneous}.
\subsection{Discretization}\label{subsec:discretization}
The numerical solution of Eq. \eqref{eq:heat_equation_incompressible_full} is obtained using the Finite Volumes (FV) method, with an implicit time scheme. The FV method is based on control volumes on which the conservation equation is integrated. There is one single unknown per control volume, the temperature, which is supposed to be an approximation of the exact solution at the center of the volume. To simplify the discretized expressions, the $N$ spatial cells have a constant space step $\Delta z$. The FV temperature in the $i$-cell is defined as:
\begin{equation}
    T_i = \frac{1}{\Delta z}\int_{z_{i-1/2}}^{z_{i+1/2}} \dd{z}T(z,t),
    \label{eq:fv}
\end{equation}
where $z_{i\pm 1/2}$ are the boundaries of the $i$-th cell. The balance equation Eq. \eqref{eq:heat_equation_incompressible_full} is divided by $\rho c$ and integrated over the $i$-th control volume. The advection term becomes:
\begin{equation}
    \frac{1}{\Delta z}\int_{z_{i-1/2}}^{z_{i+1/2}} \dd{z} \veff \pdv{T}{z} = \frac{\veff^{i+1/2} T_{i+1/2}-\veff^{i-1/2}T_{i-1/2}}{\Delta z} \simeq \frac{\veff^{i}T_{i}-\veff^{i-1} T_{i-1}}{\Delta z},
    \label{eq:advection}
\end{equation}
where the upwind scheme was used to approximate the advection term and $\veff^i = \rho_w c_w v\varphi / \rho_i c_i$. The diffusion term is discretized as:
\begin{equation}
    \frac{1}{\Delta z}\int_{z_{i-1/2}}^{z_{i+1/2}} \frac{1}{\rho c}\dd{z} \pdv{}{z}\lambda \pdv{T}{z} \simeq \frac{D_i}{\Delta z^2}\qty(\qty(T_{i+1} - T_i)\frac{\lambda_{i+1}}{\lambda_{i+1} + \lambda_{i}} - \qty(T_i - T_{i-1})\frac{\lambda_{i-1}}{\lambda_{i-1} + \lambda_{i}}),
    \label{eq:diffusion}
\end{equation}
Eq. \eqref{eq:diffusion} implicitly conserves the heat flux at the interface between two layers. Eqs. \eqref{eq:diffusion} and \eqref{eq:advection} forms a tridiagonal matrix hereafter referred as $\vb{A}$. The time derivative is discretized using an implicit scheme:
\begin{equation}
    \pdv{T}{t} \simeq \frac{\vb{T}^{(n+1)} - \vb{T}^{(n)}}{\Delta t} = \vb{A} \vb{T}^{(n+1)} + \vb{q}^{(n+1)},
    \label{eq:time}
\end{equation}
where $n$ is the time step index, $\Delta t$ is the time step and $\vb{q}^{(n)}$ is the source vector of $N$ cells containing the boundary conditions. $\vb{T}^{(n)}$ is the temperature vector at the $n$-th time step containing the $N$ FV temperatures. The implicit time scheme is chosen as it is unconditionally stable. Two non-homogeneous Dirichlet boundary conditions are used: the surface temperature is given by Eq. \eqref{eq:initial_conditions} and the deepest underground temperature is set to the annual average temperature, $C_0$. Therefore, the source vector only contains two values:
\begin{equation}
    \vb{q}^{(n+1)}_0 = T\qty(0,t^{(n+1)})\qty(\frac{D_0}{\Delta z^2} + \frac{\veff^0}{\Delta z}) \qq{and} \vb{q}^{(n+1)}_N = C_0\qty(\frac{D_N}{\Delta z^2}+\frac{\veff^N}{\Delta z}).
    \label{eq:source}
\end{equation}
Eq. \eqref{eq:time} forms a linear system of equations that must be solved at each time step knowing the previous temperature vector $T^{(n)}$. The new temperature vector is obtained by rearranging Eq. \eqref{eq:time}:
\begin{equation}
    \vb{T}^{(n+1)} = \qty(\mathbb{I} - \Delta t \vb{A})^{-1} \qty(\vb{q}^{(n+1)} + \vb{T}^{(n)}),
    \label{eq:final}
\end{equation}
where $\mathbb{I}$ is the identity matrix. The inverse of the matrix $\qty(\mathbb{I} - \Delta t \vb{A})$ is calculated using the linear algebra module of Numpy. In the remaining of this section, only the annual temperature variations are considered as their thermal skin depth is larger, Eq. \eqref{sec_roots}. 
\subsection{Results for the homogeneous medium}\label{subsec:results_homogeneous}
\begin{figure}[hbtp]
\centering
\includegraphics[width=110mm]{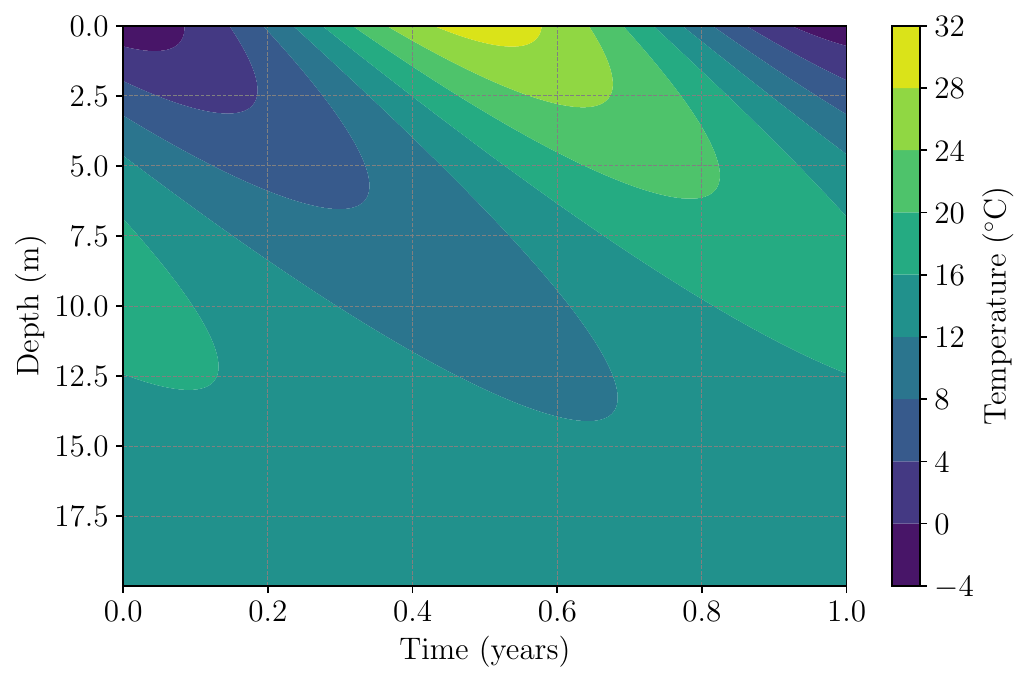}
\caption{(Color Online) Temperature evolution for annual surface variations (from January 1st to December 31), for the homogeneous medium. \(D=\SI{10e-7}{\meter^2\per\second}\) was selected, along with a water flow of \(\veff=\SI{39e-8}{\meter\per\second}\) for the calculations. Mean surface temperature is set to $\SI{13.83}{\degreeCelsius}$, with an amplitude of annual variations of $\SI{16}{\degreeCelsius}$.}
\label{fig:temperature_constant_diffusion}
\end{figure}

After an analysis of the temperature variations at Toussus-Le-Noble (a meteorological station at \SI{12}{\kilo\meter} from the location of the experiment), the coefficients of the temperature perturbation at $z=0$, Eq. \eqref{reminder_initial_conditions}, are set to $C_0 = \SI{13.83}{\degreeCelsius}$, $C_1 = \SI{0}{\degreeCelsius}$ (daily variations are ignored) and $C_2 = \SI{16}{\degreeCelsius}$. The weather data can be found on the Iowa Environmental Mesonet website \cite{herzmann2004iowa}. The space step is set to $\Delta z = \SI{0.05}{\meter}$ to account for the discontinuities of the system, Fig. \ref{coupe_sol}. The time step $\Delta t$ is set to \SI{24}{\hour}. The temperature distribution through a year is calculated up to a depth of \SI{100}{\meter}, but only the first \SI{20}{\meter} are plotted. The first calculation is run with the homogeneous model, and is presented in Fig. \ref{fig:temperature_constant_diffusion}.
As shown in Fig. \ref{fig:temperature_constant_diffusion}, the temperature fluctuations decrease with depth and are increasingly retarded in time as the depth increases. For example, at $z=\SI{10}{\meter}$, the temperature fluctuation is reduced by about a factor of $3$, and the time shift is about half a year. This agrees well with the predictions of the analytic method of Sec. \ref{sec:analytical}. A comparison between the numerical and the analytical solutions, Eq. \eqref{final_solution}, indicates a maximum difference of \SI{0.4}{\degreeCelsius}, thus validating the numerical approach.
\subsection{Results for the heterogeneous medium}\label{subsec:results_heterogeneous}
\begin{figure}[hbtp]
\centering
\includegraphics[width=110mm]{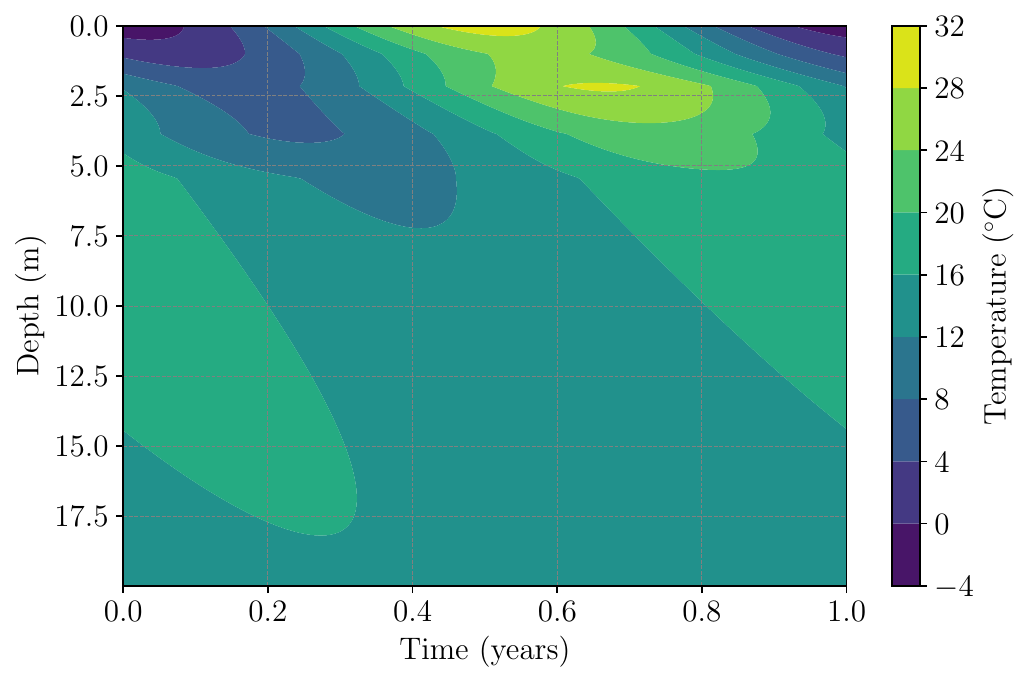}
\caption{(Color Online) Temperature evolution for annual surface variations (from January 1st to December 31), for the heterogeneous medium. The relevant thermal properties can be found in Table \ref{tab:properties}. Mean surface temperature is set to $\SI{13.83}{\degreeCelsius}$, with an amplitude of annual variations of $\SI{16}{\degreeCelsius}$.}
\label{fig:temperature_non_constant_diffusion}
\end{figure}

The parameters used for the five layers simulation presented in Fig. \ref{fig:temperature_non_constant_diffusion} are reported in Table \ref{tab:properties}. The boundary conditions applied to system are the same as for the homogeneous medium, Sec. \ref{subsec:results_homogeneous}. In addition to the two non-homogeneous Dirichlet boundary conditions, continuity of the thermal flux, Eq. \eqref{eq:fourier}, is imposed at the interfaces between through the discretized diffusion term, Eq. \eqref{eq:diffusion}. The results are presented in Fig. \ref{fig:temperature_non_constant_diffusion}.
In Fig. \ref{fig:temperature_non_constant_diffusion}, the temperature variations are also damped by the layered structure. But, the temperature distributions presented in Fig. \ref{fig:temperature_constant_diffusion} and Fig. \ref{fig:temperature_non_constant_diffusion} greatly differs. The temperature variations are much more damped in the first layers compared to the homogeneous case. This is due to the two top layers (topsoil and loess, Fig. \ref{coupe_sol}) having thermal diffusivity coefficients one order of magnitude lower than the homogenized thermal diffusivity coefficient, Table \ref{tab:properties}. Therefore, in these two layers, the perturbation Eq. \eqref{reminder_initial_conditions} is much more damped thanks to a lower thermal skin depth, Eq. \eqref{sec_roots}. A hotspot also appears at the bottom of the loess layer, which is a consequence of the advective part of Eq. \eqref{eq:heat_equation_incompressible_full}. Without the advective term, the hotspot disappears and the temperature distributions is much more damped. This underlines the capacity of rainwater to carry the temperature perturbations through the underground structure.
\section{Experimental results}\label{int:exp}
The experimental study was conducted from November 24, 2017 and March 21, 2019. Ten Pt100 thermometers, widely employed because of their precision ($\pm \SI{0.03}{\degreeCelsius}$) and ease of use, were buried in a farmer's field at the depths listed in Table \ref{tab:table_sensors}, and connected to a data logger.
\subsection{Experimental setup}\label{subsec:exp_setup}
Table \ref{tab:table_sensors} reports the depths of the sensors used in the experiment.
\begin{table}[hbtp]
\centering
\begin{tabular}{|c|c|c|c|c|c|c|c|c|c|c|}
\hline 
Sensor & 1 & 2 & 3 & 4 & 5 & 6 & 7 & 8 & 9 & 10 \\ 
\hline 
Depth (\SI{}{\centi\meter}) & 2.5 & 5 & 10 & 20 & 30 & 50 & 70 & 100 & 150 & 180 \\ 
\hline 
\end{tabular} 
\caption{Depth of the various sensors.}
\label{tab:table_sensors}
\end{table}
The spacing between thermometers increased with depth, since the amplitude of temperature variation was expected to decay exponentially with depth. Measurements were taken every 15 minutes for the entire duration of the experiment.
\subsection{Temperature vs depth}\label{subsec:temp_vs_depth}
Due to plowing after harvest season, the five sensors located less than 40 cm from the surface had to be removed during days 230-280. Even without these thermometers, the long-term variations remain clearly visible.
\begin{figure}[hbtp]
 \centering
 \includegraphics[width=140mm]{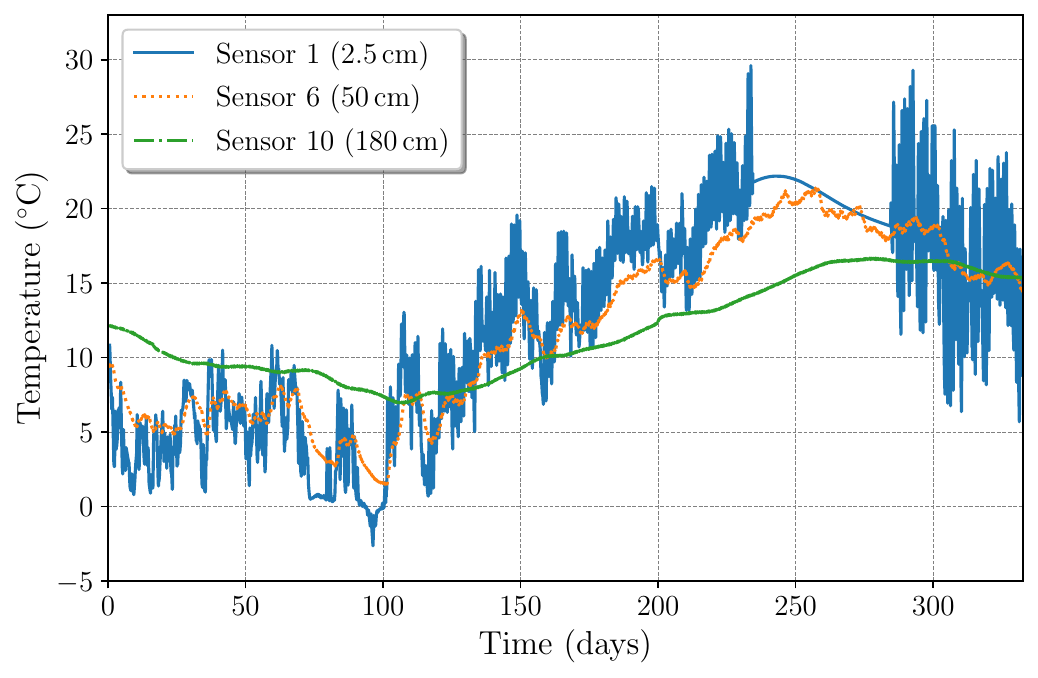}
 \caption{Example of temperature measured by three of the sensors. It can be observed that the daily temperature variations are attenuated as one goes deeper within the ground.}
 \label{ex_temp}
\end{figure}
Fig. \ref{ex_temp} shows the temperature measured by sensors located 2.5, 50, and 180 cm below the surface. Clearly, the daily temperature fluctuations decrease rapidly with depth.
\subsection{Fourier analysis}\label{subsec:fourier}
This attenuation was investigated by conducting a Fourier analysis of the data from each sensor. First, the average temperature was subtracted from the raw data (Fig. \ref{ex_temp}) for each sensor, so that each data set was centered on \SI{0}{\degreeCelsius} rather than their average temperature. Then, to prevent edge effects, a $\cos^2$ window function (or Hann window) was applied to extrapolate the data set smoothly to $0$ at both ends, effectively reducing discontinuities in the data. The interpolation is necessary as the Fast Fourier Transform (FFT) assumes periodicity of the data.  The FFT was then used to identify the dominant frequencies in the spectrum.
\begin{figure}[hbtp]
    \centering
    \includegraphics[width=140mm]{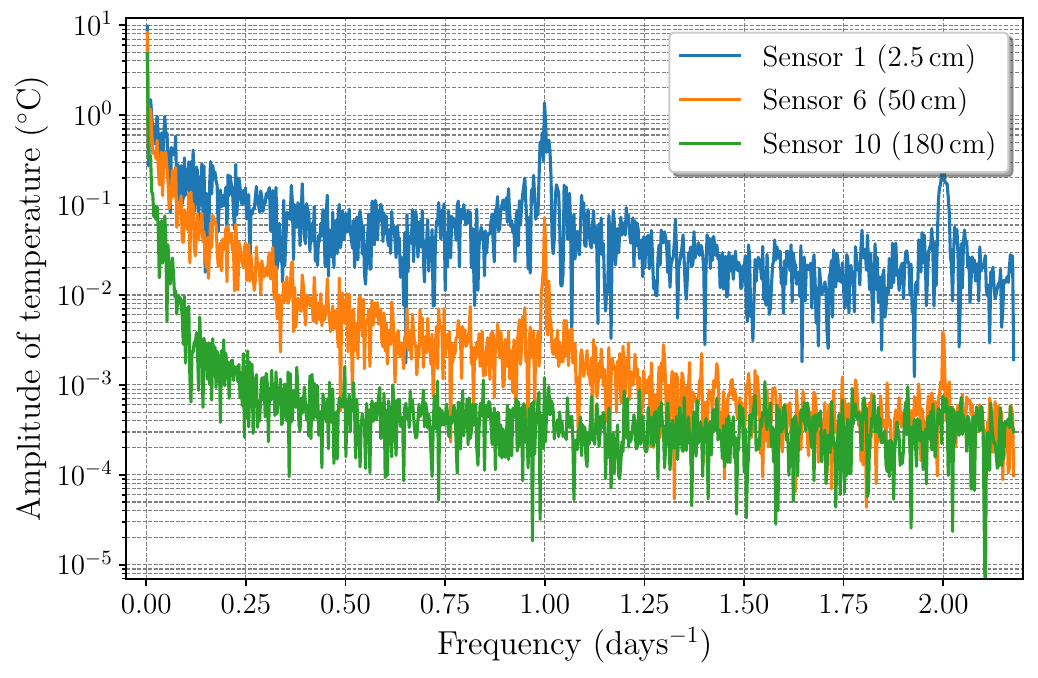}
    \caption{Spectrum of temperature amplitude. One can see that the amplitude increases as the frequency goes to zero, which corresponds to the annual variations of temperature. For sensors 1 and 6, one observes two other peaks, corresponding to a period of a day, and half a day. Note that the vertical scale is logarithmic, so the temperature variations decrease rapidly with depth.}\label{spectrum_ampl}
\end{figure}
Fig. \ref{spectrum_ampl} shows the spectra corresponding to the data of Fig. \ref{ex_temp}. For all sensors, the maximum amplitude occurs at the lowest frequency, (365 days)$^{-1}$, corresponding to the annual seasonal cycle. For all sensors, the amplitude decays rapidly with frequency, confirming that only the surface layer is affected by daily temperature variations. Sensors 1 and 6 exhibit an additional peak at a frequency (1 day)$^{-1}$, and a peak at (0.5 day)$^{-1}$ is visible from the top sensor. The daily peak corresponds to the day-night cycle, while the half-day peak is due to the daily temperature variations.
\subsection{Spectrum amplitude vs depth}\label{subsec:amplitude}
\begin{figure}[hbtp]
\centering
\includegraphics[width=140mm]{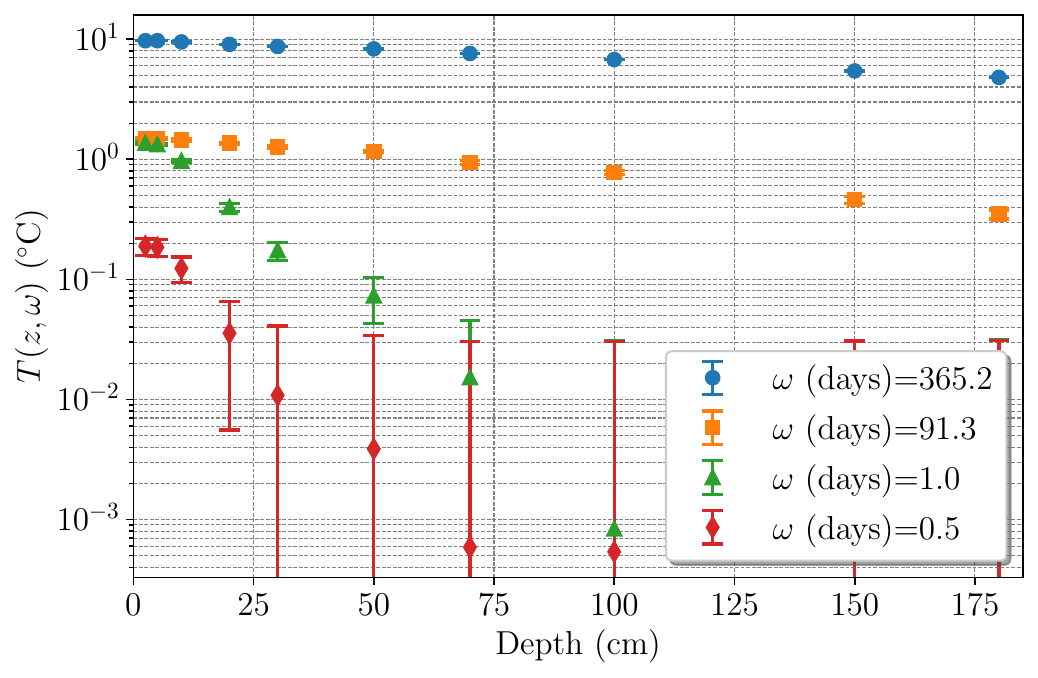}
\caption{Amplitude versus depth for different periods calculated with the FFT.}
\label{amplitude_vs_depth}
\end{figure}

Since $T(z,\omega)\propto \exp(-z / \delta (\omega))$, then $\frac{d\ln(T)}{dz} = -\frac{1}{\delta (\omega)}$. Consequently, the value of $\delta(\omega)$ can be obtained from the depth-dependent behavior of the temperature variation. Fig. \ref{amplitude_vs_depth} is a plot of the spectral amplitude $T(\omega)$ vs. depth for inverse frequencies 365, 91, 1 and 0.5 days.
As expected, the higher the frequency, the shorter the skin depth. The values of $\delta (\omega)$ inferred from Fig. \ref{amplitude_vs_depth} are listed in Table \ref{table_delta} along with the values calculated analytically (Eq. \eqref{sec_roots}).
\begin{table}[hbtp]
    \centering
    \begin{tabular}{ccc}
    \hline
    Period (days) & $\delta_{exp}$ (\SI{}{\meter}) & $\delta_{th}$ (\SI{}{\meter}) \\
    \hline
    \hline
    $365.2$ & $\SI{2.52\pm 0.04}{}$ & $\SI{6.7\pm 0.5}{}$ \\
    $91.3$ & $\SI{1.21\pm 0.04}{}$ & $\SI{2.2\pm 0.2}{} $\\
    $60.8$ & $\SI{0.96\pm 0.08}{}$ & $\SI{1.7\pm 0.1}{}$\\
    $40.5$ & $\SI{0.66\pm 0.04}{}$ & $\SI{1.3\pm 0.1}{}$\\
    $1$ & $\SI{0.17\pm 0.01}{}$ & $\SI{0.17\pm 0.02}{}$\\
    $0.5$ & $\SI{0.15\pm 0.02}{}$ & $\SI{0.12\pm 0.01}{}$\\
    \hline
    \end{tabular}
    \caption{Comparison of the experimental, theoretical and numerical values of $\delta$.}
    \label{table_delta}
\end{table}
Aside from the 365 days skin depth, the measured skin depths are in rough agreement with the theoretically calculated values, in spite of the simplicity of the experimental setup.
\section{Conclusion}
This work was undertaken to explore the potential advantages of underground housing and food storage (e.g., of wine and cheese), a strategy that has been in use since antiquity. Experimental study of the diurnal and seasonal temperature variation at the Plateau de Saclay showed that, for depths greater than \SI{2.5}{\meter}, the annual temperature variation shrinks by nearly $\SI{63}{\percent}$ ($1/e$), greatly reducing the energy required for heating and cooling.

The classical diffusion equation was introduced and extended to include advective heat flow due to rainwater circulation, and analytic solutions were found for a five-layer soil system based on the soil properties at the Plateau de Saclay, and also for a homogeneous system having properties derived by averaging those of the five layers. Numerical modeling confirmed that the heterogeneous system can be replaced by the (averaged) homogeneous one without introducing significant error. 

Pedagogically, this study introduces students to a practical and timely application of the heat equation that is highly relevant to modern geophysics and issues of importance in global warming. For instructors who may wish to assign computational projects on this topic without the need for data collection, the MIDAS Open dataset provides the underground temperature of many UK stations, with data spanning from 1900 to 2022 \cite{MetOffice2023}. This dataset is available online.
\bibliographystyle{apsrev}
\bibliography{bib.bib}
\end{document}